# Radio-Optical Alignment of Radio Galaxies

**Lizbeth Alejandra Vazquez Najar**
Universidad Autónoma de Coahuila
Facultad de Ciencias Físico Matemáticas
Unidad Camporredondo, Edificio A.
Saltillo, Coahuila, C.P. 25020, Mexico
lizbethvazqueznajar@uadec.edu.mx

**Heinz Andernach**
Universidad de Guanajuato
Departamento de Astronomía
DCNE, Sede Valenciana
Guanajuato, Gto., C.P. 36023, Mexico
heinz@astro.ugto.mx

*Resumen* — En imágenes de rastreos recientes del cielo en radioondas se midió la orientación de más de 2000 radiofuentes extendidas con galaxias huéspedes más brillantes que $m_r \sim 18$ mag. Para 1509 huéspedes se obtuvo la razón entre eje óptico mayor y menor, y para 857 de ellos también la orientación del eje mayor óptico. Se confirmó que la mayoría de las radiogalaxias eyectan sus chorros o lóbulos dentro ~30° del eje óptico menor, pero no se encontró evidencia que el alineamiento con el eje menor sea más fuerte en radiogalaxias de tamaño en radio mayor, ni tampoco para la presencia de una población secundaria de galaxias más brillantes de cúmulos con ejes en radio casi paralelos a su eje mayor óptico.

*Palabras clave* — Radiogalaxias, galaxias elípticas, rastreos del cielo en radio.

*Abstract* — Images from recent radio surveys were used to determine the radio position angles for over 2000 extended radio galaxies with host galaxies brighter than $m_r \sim 18$ mag. For 1509 of these their optical axis ratios were obtained, and for a subset of 857 also their major axis position angle. It is confirmed that the majority of radio galaxies have their jets or lobes ejected within ~30° of the optical minor axis of their host, but no evidence is found for the minor-axis alignment to be stronger for physically larger radio sources, nor for the presence of a secondary population of brightest cluster galaxies with radio axes near to parallel to their optical major axes.

## I. INTRODUCTION

Radio galaxies (RG) are almost exclusively found in elliptical galaxies and their radio emission consists of a radio nucleus (if present) at the core of the galaxy from which two oppositely directed jets are transporting relativistic particles towards diffuse outer radio lobes which often show compact emission regions (or "hotspots") indicating the place where these jets collide with the intergalactic medium. In the past, several authors have studied the relation between the optical major axis and the ejection axis of the radio jets, most of them, but not all, finding a moderate trend for these jets to be aligned to within ~30° of the optical minor axis (see Battye & Browne 2009, BB09 hereafter, and references therein). However, these samples were often limited in size and scope, and different physical characteristics were claimed by different authors as the prime cause for this alignment.

In the present work we aim to construct a sample of elliptical galaxies with extended radio emission which, in combination with previously published samples, should be large enough to be subdivided into subsamples according to various parameters. We use standard cosmological parameters: $H_0 = 70$ km s$^{-1}$ Mpc$^{-1}$, $\Omega_m = 0.3$ and $\Omega_\Lambda = 0.7$.

**II.   MATERIALS AND METHOD**

Since 2012 one of us (H.A.) has compiled a list of extended radio galaxies identified with optical or infrared (IR) objects, including their exact position, spectroscopic or photometric redshift, their brightness in optical or IR, and the angular and linear extent of their radio emission. The compilation is based on a systematic inspection of major radio survey images and on a review of over 500 publications in the literature. From over 9000 objects in the June 2019 version of this compilation we selected 2360 objects with declination > –25° and brighter than ~$18^{th}$ magnitude in r-band, such that good quality data on their optical orientation and ellipticity would be available from current optical surveys. This compilation was originally dedicated to giant radio galaxies (GRGs) with a projected linear radio size (LLS) larger than 1 Mpc, but is gradually being extended to include smaller RGs, such that our sample has a range in LLS from 10 kpc to 4.3 Mpc, with median of 440 kpc, and lower and upper quartiles of 270 and 670 kpc. The "largest angular (radio) size" (LAS) ranges from 10″ to 1° with a median of ~3′, and lower and upper quartiles of 1.9' and 5.0'. Although most quasars appear star-like on optical images, we did not exclude quasars from the beginning, so as to keep those which are hosted by galaxies of measurable optical shape.

For these 2360 objects we extracted radio images covering their full radio extent from the radio surveys NVSS (Condon et al. 1998), FIRST (Helfand et al. 2015), TGSS-ADR1 (Intema et al. 2017), and VLASS (Lacy et al. 2019). In order to recognize the structure of the radio galaxy, low- and high-resolution images of each one were displayed simultaneously with ObitView (http://www.cv.nrao.edu/~bcotton/Obit.html) and Aladin (Bonnarel et al. 2000), the latter to determine their "radio position angle" (RPA, measured from N through E). Whenever radio jets or distortions of the radio emission away from the host galaxy were present, the RPA was measured or estimated closest to the host galaxy position, and in the absence of jets the RPA was chosen along the major source axis (e.g. from one to the other hotspot). For bent-tailed sources the tangent to the jets at the host was used. This way, the RPA was measured for 2028 of the 2360 RGs by one of us (L.V.).

We then searched the *Sloan Digital Sky Survey* (SDSS DR14, Abolfathi et al. 2018) database for optical shape parameters like the optical major axis PA (or OPA in what follows, named $deVPhi_r$, $deVPhi_i$ in SDSS) and the minor-to-major axis ratio b/a ($deVAB_r$, $deVAB_i$) as well as their errors, in both the r- and i-band. The median difference between the r- and i-band OPAs was only 2°, so we used the r-band OPA whenever this difference was below 10°, and whenever the relative error of the r-band axis ratio was less than 0.2. Otherwise we ignored the OPA. We also excluded the shape parameters of all objects classified photometrically as "star", i.e. quasars with no detected host galaxies around them. This resulted in 1453 SDSS objects with decent axis ratios, but, following BB09, we considered their OPAs only for b/a < 0.8.

For the objects not covered by SDSS we then searched the VizieR catalog browser (Ochsenbein et al. 2000) for matches with the SkyMapper DR1 survey (Wolf et al. 2018) which offers shape parameters for galaxies.  For 181 objects with Decl < 0° we found 172 matches, of which we excluded those with stellarity class >0.65 and major axis <3 pixels, for being too compact. This left 79 objects with axis ratios, for which we kept OPAs only for seven objects which had both, an error in OPA of <10° and an axis ratio of b/a < 0.8. We further obtained optical shapes for 106 objects from the DECaLS survey DR7 (Dey et al. 2019), but we accepted OPAs only for 59 objects with b/a < 0.8. For the 67 DECaLS objects in common with SDSS we found good agreement in the shape parameters. For the remaining 387 objects with measured RPAs but no optical shape parameters (excluding those we already rejected from SDSS), we searched HyperLEDA (Makarov et al. 2014) within a radius of 3″, finding 244 matches. We used 173 of these which had axis ratios with errors less than 0.2 and accepted their OPAs for 104 objects with b/a < 0.8.

### III. RESULTS AND ANALYSIS

For the sample of 2028 RGs for which we had measured RPAs, our search for optical shape parameters resulted in 1509 objects with available axis ratio b/a. For 857 of these we also had their OPAs and calculated the acute difference angle dPA = | RPA – OPA |, such that for dPA=90° the radio axis is aligned with the optical minor axis and for dPA=0° with the major axis. Spectroscopic redshifts are known for 85 % of the 1509 RGs, and for 87 % of those with both RPA and OPA, and for all others reliable photometric redshifts were found in various references, like Brescia et al. (2014), Bilicki et al. (2014, 2016) and others. The median redshift is 0.12 and the maximum is 0.38. All except 11 objects have Galactic latitude |b| > 10°, and 134 are GRGs (LLS > 1 Mpc), of which 92 have both RPA and OPA.

We confirm the trend for minor-axis alignment (Fig. 1), as well as the trend reported by BB09 for the minor-axis alignment to strengthen for rounder (larger b/a) galaxies (Fig. 2). However, while BB09 used a larger sample of ~5000 early-type galaxies (their figures 5 and 7, right panels) these authors did not publish their sample, nor did they include parameters like LLS, redshift, radio morphology and radio luminosity, or the host's rank in a galaxy cluster. Here we try to consider the possible relevance of these parameters.

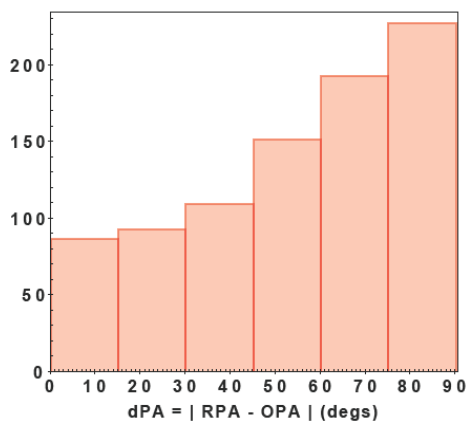

**Fig. 1.** Distribution of the radio-optical misalignment angle for all 857 objects.

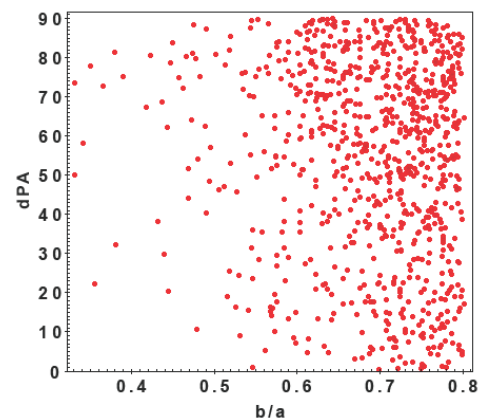

**Fig. 2.** Radio-optical misalignment angle (°) as function of the optical axis ratio.

Palimaka et al. (1979) reported a trend for the minor-axis alignment to become stronger with the LLS of the radio sources. Indeed, figure 3 shows that all but one of the GRGs with LLS > 2 Mpc have a dPA > 50°. However, a Kolmogorov-Smirnov (KS) test comparing the dPA distribution of 200 RGs with LLS > 700 kpc with that of 210 RGs with LLS < 250 kpc shows no evidence for any difference in dPA ($p_{KS}$ = 0.23). For those 652 objects of our sample for which the radio-morphological FR type (Fanaroff & Riley 1974) was well determined, we compared the 195 RGs with of type FR I with 457 of type FR II, and found a median dPA of 54.9° and 63.6° for FR Is and FR IIs, respectively, with a high significance ($p_{KS}$= 0.002) of being drawn from different populations. However, a decision on whether this difference is primarily due to FR type or radio luminosity (see figure 4 in Best 2009), requires our determination of the latter.

Separating the 857 radio galaxies with dPA values at their median redshift of 0.132, we find no evidence for any difference in the distribution of dPA for the low- and high-redshift halves of the sample ($p_{KS}$=0.16). Likewise, the 28 quasars do not show any difference in their dPA distribution from the 829 galaxies, but the axis ratio b/a is higher for the quasars at the $p_{KS}$=0.018 level, most likely because their bright optical nucleus makes them appear rounder on optical images. Finally, no significant trend is seen between the linear radio extent (LLS) and the axis ratio (b/a) of their host galaxies.

Brightest cluster galaxies (BrClGs) have been claimed to have a more bimodal distribution of their dPA with a secondary, smaller maximum at dPA=0° (Andernach & Ramos Ceja 2009). Our present sample was assembled independent of the latter one but does contain (albeit incomplete) tags indicating whether an RG host is also a BrClG. Of the 114 RGs tagged as BrClGs in our present sample we find the dPA distribution shown in figure 4. Despite the impression of a marginal peak at dPA=0°, neither the KS nor the Mann-Whitney U test provide any evidence for the BrClG and non-BrClG samples being drawn from different populations.

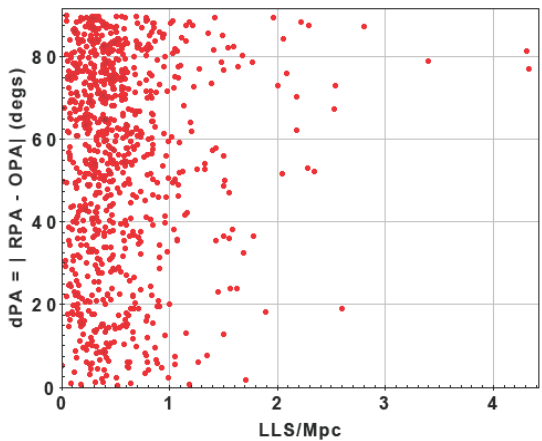

**Fig. 3.** Radio-optical difference angle dPA vs. largest linear size for 857 RGs.

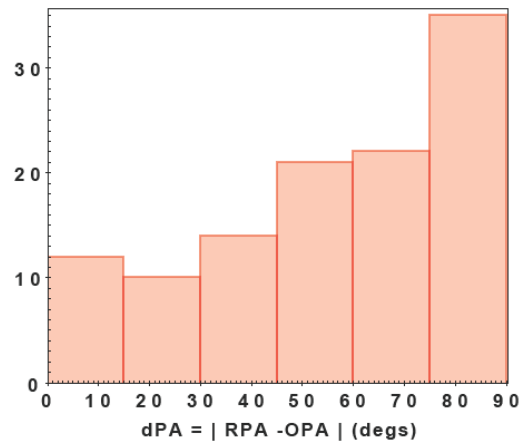

**Fig. 4.** Distribution of dPA for 114 RGs tagged as brightest cluster galaxies (BrClG).

## IV. CONCLUSIONS AND FUTURE WORK

We used recent radio survey images to measure the radio position angles for over 2000 extended radio galaxies with optical host galaxies brighter than 18$^{th}$ magnitude in r band. From available optical surveys and databases we extracted the optical axis ratio for 1509 of their host galaxies, as well as the major axis position angle for a subset of 857. We confirm the well-known trend that the majority of radio galaxies have their jets or lobes aligned to within ~30° (dPA > 60°) of the optical minor axis of their host, but do not find evidence for previous claims that (a) the minor-axis alignment is stronger for physically larger radio sources or (b) there is a secondary population of brightest cluster galaxies having their radio axes aligned with their optical major axes. However, the distributions of dPA for FR I and FR II sources show a stronger tendency for minor-axis alignment for the FR IIs. Given that FR IIs tend to be more radio luminous, this is in apparent contrast with BB09 (their figure 9) who found a stronger minor-axis alignment for radio-louder objects, albeit based on only the FIRST core fluxes which may underestimate the total flux for a large fraction of sources. We plan to determine both optical and radio luminosities, the latter by flux integration on survey images, which should reveal whether the dPA distribution depends more directly on radio luminosity or FR type.

The seven weeks of this project were not enough to check the reliability of the optical shapes for every galaxy or quasar host. Some of these will be affected by superposed Galactic stars or other galaxies, as well as suffer from isophote twist (see e.g. Porter et al. 1991), such that a distinction will have to be made between inner and outer optical position angle. Moreover, optical shapes for many other host galaxies may be obtained from PanSTARRS1 images (Flewelling et al. 2016), and more brightest cluster galaxies may be identified from large cluster catalogues (e.g. Wen et al. 2018 and references therein).

The current selection of objects is biased towards rather large radio sources, and we shall work towards including previously published samples with known RPA and OPA as well as physically smaller sources hosted by any galaxy brighter than $m_r$ ~ 18 mag, like most of those used by BB09. Where possible, we shall derive integrated radio spectral indices, to see whether they have any influence on the radio-optical alignment.

## V. ACKNOWLEDGEMENTS

A significant number of RGs in our sample was found by volunteers of the Radio Galaxy Zoo project (http://rgzauthors.galaxyzoo.org). We are grateful to the VLASS team at NRAO for a timely delivery of Quick Look images, and to F.J. Peralta for his python script to cut out images from the VLASS survey. Jean Tate helped with the extraction of optical host parameters, and H.A. benefited from research grant CIIC 218/2019 of Univ. of Guanajuato. We made use of NASA's SkyView facility (http://skyview.gsfc.nasa.gov ) located at NASA Goddard Space Flight Center.